\documentclass[12pt,preprint]{aastex}
\usepackage{bm}
\usepackage{threeparttable}
\usepackage{amsmath,amssymb}
\usepackage{textcomp}
\usepackage{booktabs}
\usepackage{lscape}
\makeatletter
\newcommand{\tblcaption}[1]{\def\@captype{table}\caption{#1}}
\makeatother
\usepackage{color}

\title{MOA-2012-BLG-505Lb: A super-Earth mass planet probably in the Galactic bulge}

\author{M. Nagakane\altaffilmark{1}, T. Sumi\altaffilmark{1}, N. Koshimoto\altaffilmark{1}, D. P. Bennett\altaffilmark{2,3}, I. A. Bond\altaffilmark{4}, N. Rattenbury\altaffilmark{5}, D. Suzuki\altaffilmark{2}}

\and

\author{F. Abe\altaffilmark{6}, Y. Asakura\altaffilmark{6}, R. Barry\altaffilmark{2}, A. Bhattacharya\altaffilmark{2,3}, M. Donachie\altaffilmark{5}, A. Fukui\altaffilmark{7}, Y. Hirao\altaffilmark{1}, Y. Itow\altaffilmark{6}, M. C. A. Li\altaffilmark{5}, C. H. Ling\altaffilmark{4}, K. Masuda\altaffilmark{6}, Y. Matsubara\altaffilmark{6}, T. Matsuo\altaffilmark{1}, Y. Muraki\altaffilmark{6}, K. Ohnishi\altaffilmark{9}, C. Ranc\altaffilmark{2}, To. Saito\altaffilmark{10}, A. Sharan\altaffilmark{5}, H. Shibai\altaffilmark{1}, D. J. Sullivan\altaffilmark{11}, P. J. Tristram\altaffilmark{12}, T. Yamada\altaffilmark{1}, A. Yonehara\altaffilmark{13}\\ (MOA Collaboration)}

\altaffiltext{1}{Department of Earth and Space Science, Graduate School of Science, Osaka University, 1-1 Machikaneyama, Toyonaka, Osaka 560-0043, Japan}
\altaffiltext{2}{Laboratory for Exoplanets and Stellar Astrophysics, NASA/Goddard Space Flight Center, Greenbelt, MD 20771, USA}
\altaffiltext{3}{Department of Physics, University of Notre Dame, Notre Dame, IN 46556, USA}
\altaffiltext{4}{Institute of Information and Mathematical Sciences, Massey University, Private Bag 102-904, North Shore Mail Centre, Auckland, New Zealand}
\altaffiltext{5}{Department of Physics, University of Auckland, Private Bag 92019, Auckland, New Zealand}
\altaffiltext{6}{Institute for Space-Earth Environmental Research, Nagoya University, Nagoya, 464-8601, Japan}
\altaffiltext{7}{Okayama Astrophysical Observatory, National Astronomical Observatory, 3037-5 Honjo, Kamogata, Asakuchi, Okayama 719-0232, Japan}
\altaffiltext{8}{European Southern Observatory (ESO), Karl-Schwarzschildst. 2, D-85748 Garching, Germany}
\altaffiltext{9}{Nagano National College of Technology, Nagano 381-8550, Japan}
\altaffiltext{10}{Tokyo Metropolitan College of Industrial Technology, Tokyo 116-8523, Japan}
\altaffiltext{11}{School of Chemical and Physical Sciences, Victoria University, Wellington, New Zealand}
\altaffiltext{12}{University of Canterbury Mt. John Observatory, P.O. Box 56, Lake Tekapo 8770, New Zealand}
\altaffiltext{13}{Department of Physics, Faculty of Science, Kyoto Sangyo University, Kyoto 603-8555, Japan}


\newpage

\begin{document}

\begin{abstract}

We report the discovery of a super-Earth mass planet in the microlensing event MOA-2012-BLG-505.
This event has the second shortest event timescale of $t_{\rm E}=10 \pm 1$ days where the observed data show evidence of planetary companion. 
Our 15 minute high cadence survey observation schedule revealed the short subtle planetary signature. 
The system shows the well known close/wide degeneracy. 
The planet/host-star mass ratio is $q =2.1 \times 10^{-4}$ and the projected separation normalized by the Einstein radius is s = 1.1 or 0.9 for the wide and close solutions, respectively. 
We estimate the physical parameters of the system by using a Bayesian analysis and find that the lens consists of a super-Earth with a mass of $6.7^{+10.7}_{-3.6}M_{\oplus}$ orbiting around a brown-dwarf or late M-dwarf host with a mass of $0.10^{+0.16}_{-0.05}M_{\odot}$ with a projected star-planet separation of $0.9^{+0.3}_{-0.2}$AU. 
The system is at a distance of $7.2 \pm 1.1$ kpc, i.e., it is likely to be in the Galactic bulge.
The small angular Einstein radius ($\theta_{\rm E}=0.12 \pm 0.02$ mas) and short event timescale are typical for a low-mass lens in the Galactic bulge. 
Such low-mass planetary systems in the Bulge are rare because the detection efficiency of planets in short microlensing events is relatively low. 
This discovery may suggest that such low mass planetary systems are abundant in the Bulge and currently on-going high cadence survey programs will detect more such events and may reveal an abundance of such planetary systems.

\end{abstract}

\maketitle
\newpage

\section{Introduction}
\label{sec-intro}

Since the first discovery of an exoplanet orbiting a main sequence star \citep{1995Natur.378..355M}, more than 3000 exoplanets have been discovered.
Most of these were discovered by the radial velocity method \citep{2006ApJ...646..505B} and the transit method \citep{2011ApJ...736...19B}.
Although these methods have a higher sensitivity to planets with higher masses and closer orbits in general, their sensitivities are evolving to allow the discovery of planets of lower mass at wider orbits. 
The Kepler satellite revolutionized our understanding of the exoplanet distribution by detecting many small planets down to Earth-radius planets with semi major axes of less than 1 AU. 
However the population of low-mass planets with separations of a few AU are less understood. \\

Exoplanet searches using gravitational microlensing were first proposed by \citet{1991ApJ...374L..37M}.
If a background source star is closely aligned with a foreground lens star, the gravity of the lens bends the light from the source star to create unresolvable images of the source, yielding an apparent magnification of the source star brightness. 
If the lens star has a planetary companion and it lies close to one of the source images, the gravity of the planet perturbs the observed light curve.
Microlensing is uniquely sensitive to exoplanets at orbit radii 1-6 AU, just outside of the snow-line \citep{2004ApJ...616..567I, 2004ApJ...612L..73L, 2006ApJ...650L.139K} with masses down to Earth mass planets.
Contrary to the other planet detection methods mentioned above, microlensing does not rely on any light from the host star. 
Microlensing is also sensitive to low-mass planets \citep{1996ApJ...472..660B} orbiting the faint and/or distant stars, like late M-dwarfs, the most common stars in our Galaxy. \\

Since the first discovery of exoplanets by microlensing \citep{2004ApJ...606L.155B}, about 50 planetary microlensing events have been found.
Many microlensing planets have been discovered in high-magnification ($A \gtrsim 100$) events.
Although the high magnification events are rare, the events are very sensitive to planets \citep{1998ApJ...500...37G, 2000ApJ...533..378R, 2002MNRAS.335..159R}. 
Therefore, we issue alerts for high magnification events to encourage follow-up observations to detect and characterize planetary anomalies in the light curves.
We report the discovery of a planetary system in the Galactic bulge in the high amplification microlensing event MOA-2012-BLG-505. \\

We describe the observations and the data set of this event in Section \ref{sec-obs}. 
The light curve modeling is described in Section \ref{sec-model}. 
Section \ref{sec-cmd} presents the data calibration and the source star radius estimate. 
In Section \ref{sec-bayes}, we present a Bayesian analysis which is used to estimate the physical parameters of the lens system. 
Finally, our discussion and conclusions are given in Section \ref{sec-discuss}.

\section{Observation and Data}
\label{sec-obs}

The Microlensing Observations in Astrophysics (MOA; \citealp{2001MNRAS.327..868B,2003ApJ...591..204S}) collaboration is conducting a microlensing exoplanet search towards the Galactic bulge using the 1.8m MOA-II telescope at Mt. John University Observatory (MJUO) in New Zealand.
Thanks to the wide field of view (FOV) of 2.2 $\rm deg^{2}$ with $\rm 10k \times 8k$ pixel mosaic CCD-camera, MOA-cam3 \citep{2008ExA....22...51S}, we conduct a high cadence survey observations depending on the field.
The observations are carried out mainly with a custom broad $R+I$ band filter, called MOA-Red. \\

On 2012 July 27, UT 17:48, the microlensing event MOA-2012-BLG-505 was detected and alerted by MOA at $(\alpha, \delta)(2000) = (17^{h}52^{m}34^{s}.34,-32^{\circ}02^{\prime}24^{\prime \prime}.33)$ corresponding to Galactic coordinates: $(l, b) = (-1.892^{\circ}, -2.881^{\circ})$.
This event was alerted at HJD - 2450000 = 6136.24 and, because the event was very short and fast-rising, the alert was issued after peak magnification and when the planetary signal at the peak was already over.
Therefore, no follow-up observation was conducted on this event by other groups. 
Fortunately, this event occurred in one of fields with the highest cadence observation of 15min.
Thus we could detect a short subtle planetary signature by MOA data alone. 
Figure 1 shows the observed light curve. \\

The MOA data were reduced with MOA's implementation of a Difference Image Analysis (DIA) pipeline \citep{2001MNRAS.327..868B}. 
This method has an advantage for detecting objects whose brightness changed in the stellar crowded fields like those in the Galactic bulge by subtracting good-quality reference images from each of the observed images.
It also produces better photometric light curves, avoiding the effect of blending stars compared to traditional PSF photometry. 
Moreover, our pipeline is designed to be sensitive to those events whose source star is fainter and not resolved in the reference image, but magnified brighter than observational limiting magnitude by the microlensing, as in this event. \\

It is known that the nominal error bars calculated by the pipeline are misestimated in such stellar crowded fields due to various reasons.
We empirically renormalized the errors by using the standard method \citep{2012ApJ...755..102Y} as follows.
Note that this renormalization is intended to get proper errors of the parameters in the light curve modeling, and we made sure that the fitting results does not depend on this renormalization.
We renormalized by using the formula,
\begin{equation}
  \sigma^{\prime}_{i} = k \sqrt{\sigma^{2}_{i} + e^{2}_{min}}
\end{equation}
where $\sigma^{\prime}_{i}$ is the $i$th renormalized error, $\sigma_{i}$ is the $i$th error obtained from DIA, and $k$ and $e_{min}$ are the renormalizing parameter. \\

At first, we searched the preliminary best fit model by using light curves with original errors.
Then we sort the data points by magnification and make a cumulative $\chi^{2}$ distribution. 
The $e_{min}$ value is chosen so as to the slope of  distribution to be 1. 
The $k$ value is chosen so as to $\chi^{2}/dof \simeq 1$. 
As a result, we obtained $e_{min} = 0$ and $k = 1.351988$ for MOA-Red, $e_{min} = 0$ and $k = 0.932111$ for MOA-V. \\

\section{Light Curve Models}
\label{sec-model}

There are five microlensing parameters for a point-source point-lens (PSPL) model: the time of lens-source closest approach $t_{0}$, the Einstein timescale $t_{\rm E}$, the minimum impact parameter in units of the Einstein radius $u_{0}$, the source flux $f_{\rm S}$ and the blend flux $f_{\rm B}$. 
There are three more parameters for a point-source binary (planetary)-lens model: the planet-host mass ratio $q$, the planet-host separation in units of the Einstein radius $s$, the angle between the trajectory of the source and the planet-host axis $\alpha$.
Moreover, if the finite size of the source is considered (finite source effect), we need the source size in units of the Einstein radius $\rho \equiv \theta_{\rm *}/\theta_{\rm E}$ where $\theta_{\rm *}$ is the angular source radius, $\theta_{\rm E}$ is the angular Einstein radius.
If the effect of Earth's orbital motion during the event, called the microlensing parallax effect, is significant, the north and east components of the microlensing parallax vector $\pi_{\rm E,N}$ and $\pi_{\rm E,E}$ are added, respectively. 
If both the finite source effect and the microlensing parallax effect are detected, we can determine the lens mass and the distance directly \citep{2011ApJ...741...22M}. \\

To fit the microlensing parameters, we used the Markov Chain Monte Carlo (MCMC) algorithm \citep{2003ApJS..148..195V} and the \citet{2010ApJ...710.1641S} implementation of the image centered ray-shooting method \citep{1996ApJ...472..660B, 2010ApJ...716.1408B}. 
Three microlensing parameters, $(q,s,\alpha)$, feature the anomaly shape well, thus we first conduct a grid search of 9680 fixed grid points of these parameters leaving the other parameters free. 
We used 11 values for $\log q$ over the range $\log q \in [-4,0]$, 22 values for $\log s$ in the range $\log s \in [-0.5,0.55]$ and 40 values of $\alpha$ in the range $\alpha \in [0,2\pi]$.
Then we searched for the best model by refining the best 100 models from the initial grid search with all parameters free. 
In this event, we detected a finite source effect. 
To obtain the value of $\rho$ properly, we included linear limb-darkening. 
The stellar effective temperature $T_{\rm eff}$ computed from the source color presented in Section \ref{sec-cmd} is $T_{\rm eff} \sim 6213 \, \rm K$ \citep{2009A&A...497..497G}. 
We assumed $T_{\rm eff} \sim 6250 \, \rm K$, a surface gravity of $\log g = 4.5 \, \rm cm \, s^{-2}$, the microturbulent velocity as $v_{\rm t} = 0 \, \rm km \, s^{-1}$, and a metalicity of $\log [M/H] = 0$. 
We used the corresponding limb-darkening coefficient from the ATLAS stellar atmosphere models of \citet{2000A&A...363.1081C}, and found the coefficient for MOA-Red is $u=0.52845$ by taking the mean of the R and I-band values. 
We also determined the coefficient for MOA-V is $u=0.6413$. \\

Because the peak magnification of the event is $A_{\rm max} \sim 100$, the event is very sensitive to the planetary perturbations.
Figure 1 shows the best-fit model and the light curve of this event. 
Table 1 presents the best-fit model parameters.
We found that the best-fit model with planetary mass ratio of $q=2 \times 10^{-4}$ reproduces the asymmetric feature at the peak of light curve in this event. 
Comparing to the single lens model, the best planetary model improves $\chi^{2}$ by $\Delta \chi^{2} \sim 227$. 
Therefore, the planetary signal is detected significantly. \\

The inclusion of the finite source effect improves the fit by $\Delta \chi^{2} = 17.1$, i.e., larger than $4\sigma$. 
The comparison between the model with and without the finite source is shown in Figure 2.
One can see that the model with the finite source fits the data better around the event peak. 
Thus, we include the finite source effect in the following analysis. 
Because the source trajectory of the best-fit model crosses only the central caustic as is common in high magnification events, there is the well known “close-wide” degeneracy. 
The microlensing parameters are almost same except for the separations. 
The models with $s<1$ and $s>1$ are called the “close” and “wide” models, respectively. 
In this event, the shapes of the central caustics for each close and wide models are fairly similar as shown in Figure 3. 
The $\chi^{2}$ difference between these models is only $\Delta \chi^{2} = 0.019$, therefore we cannot distinguish them. \\

We did not detect any the microlensing parallax effect in this event. 
This is as expected because the event is short with timescale of $\sim 10$ days \citep{1992ApJ...392..442G, 2012ARA&A..50..411G}. \\

We also found a model with a stellar mass ratio of $q \sim 6.7 \times 10^{-2}$, which is the second preferred model with $\Delta \chi^{2} \sim 32$ compared to the best close-wide planetary models above.
Although the best planetary models have significant $\Delta \chi^{2}$ compared to the best single lens model, the distinction of these models against stellar binary models is marginal because of relatively subtle anomaly features. 
To confirm that our planetary model best explains the observed data, we conducted a further grid search to check if there are any other local minima with different $q$ values. 
In Figure 4, we plot $\Delta \chi^{2}$ for each model with a range of mass ratios $q$. 
Here, we searched for the best model using each $q$ value leaving the other parameters to vary freely. 
The ranges of $q$, $s$ and $\alpha$ in this search are same as those used in the first grid search. 
We found some local minima corresponding to distinctly different lens system geometries. 
Three of these minima can be seen in Figure 4 while the others have larger $\chi^{2}$. 
Models in the range of $q \lesssim 4 \times 10^{-3}$, $4 \times 10^{-3} \lesssim q \lesssim 1 \times 10^{-2}$ and $q \gtrsim 1 \times 10^{-2}$ have the same sort of geometry as the best model with $(q, s, \alpha) = (2.1 \times 10^{-4}, 1.1, 2.4)$, another planetary solution with $(q, s, \alpha) = (5.2 \times 10^{-3}, 1.3, 5.0)$ and the stellar binary solution with $(q, s, \alpha) = (6.6 \times 10^{-2}, 0.2, 1.3)$, respectively. 
We found that the local minimum with $q \sim 6.63 \times 10^{-2}$ in Figure 4 is the only local minimum having a the stellar mass ratio. 
This model corresponds to the second best model found from the first grid analysis above whose $\Delta \chi^{2}$ with respect to the best model is $\sim 31$ (i.e., larger than $5\sigma$). 
Therefore, we conclude that the planetary solution is the global best solution for this event. \\

Because we were unable to detect microlensing parallax in this event, we cannot determine the physical parameters of the lens system uniquely. 
Only the measurement of the finite source effect in this event partially breaks the degeneracy between microlensing parameters and that can be used to estimate the probability distribution of the physical parameters via the Bayesian analysis in Section \ref{sec-bayes}. \\

\section{CMD and Source Radius}
\label{sec-cmd}

In this section we estimate the angular Einstein radius $\theta_{\rm E}=\theta_{\rm *}/\rho$ from the best fit $\rho$ and the angular source radius $\theta_{\rm *}$ which can be calculated from the source color and magnitude. 
We get the source color and magnitude from MOA-Red band and MOA-V band data. 
Figure 5 shows the OGLE (the Optical Gravitational Lensing Experiment; \citealp{2003AcA....53..291U}) $(V-I,I)$ color-magnitude diagram (CMD) of stars within $2^{\prime}$ around MOA-2012-BLG-505 \citep{2011AcA....61...83S}. 
It also shows the deep CMD of Baade's window as observed by HST \citep{1998AJ....115.1946H} and which is adjusted for the distance, reddening and extinction to the MOA-2012-BLG-505 field by using Red Clump Giants (RCG) as standard candles \citep{2008ApJ...684..663B}. 
We convert the best fit MOA-Red and MOA-V source magnitude to the standard Cousins I and Johnson V magnitude by cross-referencing stars in the MOA field with stars in the OGLE-III photometry map \citep{2011AcA....61...83S} within $2^{\prime}$ of the event. 
We find the source color and magnitude to be $(V-I,I)_{\rm S,OGLE} = (2.00, 21.29) \pm (0.06, 0.13)$. 
We only have a couple of MOA-V data points, one highly magnified point and one with low magnification, during this event. 
So we doubled the nominal error of the source color conservatively. 
We find, therefore, the source color and magnitude to be $(V-I,I)_{\rm S,OGLE} = (2.00, 21.29) \pm (0.12, 0.13)$. 
The centroid of RCG color and magnitude in CMD are $(V-I,I)_{\rm RCG} = (2.49, 16.32) \pm (0.01, 0.03)$ as shown in Figure 5. 
Comparing these values to the expected extinction-free RCG color and magnitude at this field of $(V-I,I)_{\rm RCG,0} = (1.06, 14.55) \pm (0.07, 0.04)$ \citep{2013A&A...549A.147B, 2013ApJ...769...88N}, we get the reddening and extinction to the source of $(E(V-I),A_{I})_{\rm RCG} = (1.43, 1.77) \pm (0.07, 0.05)$. 
Therefore, we estimated the extinction-free source color and magnitude as being
\begin{equation}
  (V-I,I)_{\rm S,0} = (0.57, 19.52) \pm (0.14, 0.14).
\end{equation}
By using the empirical formula, $\log(\theta_{\rm *}) = 0.50141358 + 0.41968496(V-I) - 0.2I$ \citep{2014AJ....147...47B, 2015ApJ...809...74F}, we estimated the angular source radius, 
\begin{equation}
  \theta_{\rm *} = 0.34 \pm 0.05 \ {\rm \mu as}.
\end{equation}
From this $\theta_{\rm *}$ and other fitting parameters, we calculated the angular Einstein radius $\theta_{\rm E}$ and the lens-source relative proper motion $\mu_{\rm rel} = \theta_{\rm E} / t_{\rm E}$, as follows, 
\begin{eqnarray}
  \theta_{\rm E} & = & 0.12 \pm 0.02 \ {\rm mas} \\
  \mu_{\rm rel} & = & 4.40 \pm 0.72 \ {\rm mas \ yr^{-1}} 
\end{eqnarray}
This $\theta_{\rm E}$ is relatively small which indicates that lens is likely low-mass and/or far from the observer. 
This $\mu_{\rm rel}$ is relatively high which prefers that the lens is in the Galactic bulge rather than the disk.

The source color obtained from MOA-Red and MOA-V data lies on the blue edge of the typical main sequence stars expected from the HST CMD. 
We checked the effect of possible systematics in the color measurement on the final estimated lens physical parameters as follows. 
Assuming the source star is a main sequence star in the Bulge, which is very likely, we adopt the mean color of the main sequence stars in the HST CMD with the magnitudes ranging around the observed source magnitude $I_{S} = 21.29 \pm 0.13$ \citep{2008ApJ...684..663B}.
We get a source color and magnitude of $(V-I,I)_{\rm S,OGLE} = (2.32, 21.29) \pm (0.12, 0.13)$ and $(V-I,I)_{\rm S,0} = (0.89, 19.52) \pm (0.14, 0.14)$, with and without extinction and reddening, respectively.

The lens physical parameters estimated by using this source color are consistent with the final results by using the source color obtained from the light curves (see Section \ref{sec-bayes}) to within 1$\sigma$ errors. 
Therefore, systematics in our measurement of the source color do not make any difference to our final results. 
We use the source color measured from the light curves in the following discussions.

\section{Lens Physical Parameters by Bayesian Analysis}
\label{sec-bayes}

We can not directly determine the lens physical parameters because we couldn't measure the microlensing parallax effect. 
Thus we estimated the probability distribution of lens physical properties by using a Bayesian analysis \citep{2006Natur.439..437B, 2006ApJ...644L..37G, 2008ApJ...684..663B} assuming the Galactic model of \citet{1995ApJ...447...53H} as a prior probability. 
We note that the probability of stars hosting a planet with the measured $q$ and $s$ values is independent of the lens's mass and distance from the Earth. 
We used the measured $\theta_{\rm E}$ and $t_{\rm E}$ to constrain the probability distributions of lens parameters. 
Although this event has the $s \leftrightarrow 1/s$ degeneracy, all lens physical parameters of these degenerate solutions are consistent within the $68.3\%$ confidence interval.
Therefore, we combined the result of wide and close model weighted by $e^{-\Delta \chi^{2}/2}$ ($\Delta \chi^{2} = \chi^{2}_{close}-\chi^{2}_{wide} = 0.019$). \\

Planetary system parameters confidence intervals are sometimes set at the fixed contour levels, but this is confusing for distributions with multiple peaks. 
So, we prefer percentiles, so that the 68$\%$ and 95$\%$ confidence intervals represent the central 68$\%$ and 95$\%$ of the probability distribution. 
The posterior distributions and parameters are shown in Figure 6, Figure 7 and Table 2, respectively.
According to this analysis, the host is a brown-dwarf or late M-dwarf with a mass of $0.10^{+0.16}_{-0.05}M_{\rm \odot}$ at $7.21^{+1.14}_{-1.11}$kpc away from the Earth i.e., likely in the Galactic bulge. 
In the calculation of the posterior probability, we treated the stars in the Bulge and the stars in the disk separately following each number density distributions given by \citet{1995ApJ...447...53H}. 
The posterior probability that the lens system is in the bulge is $80.5 \%$, and the probability that the lens primary is a brown dwarf in the disk is $19.5 \%$. 
The planet mass is a super-Earth with a mass of $6.7^{+10.7}_{-3.6}M_{\rm \oplus}$ and the projected separation from the host is $r_{\rm \perp} = 0.91^{+0.26}_{-0.23}$AU. 
Assuming a planetary orbit with random inclination and phase \citep{1992ApJ...396..104G}, the physical three-dimensional separation is $a = 1.13^{+0.68}_{-0.34}$AU. \\

\section{Discussion and Conclusion}
\label{sec-discuss}

We found that MOA-2012-BLG-505Lb is a super-Earth mass planet, ranging from terrestrial mass to Neptune mass, orbiting around a brown-dwarf or late M-dwarf probably in the Galactic bulge. 
Figure 8 shows the distribution of known exoplanets in planet mass as a function of the host mass. 
This event is shown by a purple circle and the position is bottom-left (corresponding to both a low-mass host and planet). 
The red circles indicate the planets found by microlensing, where filled and open circles indicates that their masses are measured and estimated by a Bayesian analysis, respectively. 
Red ones indicate planetary mass ratios of $q \leq 0.1$ and green ones, on the other hand, indicate larger mass ratios of $q > 0.1$, i.e., low mass binaries. 
Although the detection efficiencies of the planetary events with large $q$ are higher than that with small $q$ \citep{2016ApJ...833..145S}, all of planetary systems have small $q \sim 0.001-0.0001$ except one planetary system with large $q \sim 0.01$ \citep{2016AJ....152...95H} orbiting around low-mass stars of $< 0.15 M_{\odot}$.
This result indicates that the low-mass stars tend to host relatively low-mass planetary companions. \\

This event is roughly tied for the second shortest $t_{\rm E}\approx 10\,$days with a planetary companion with OGLE-2016-BLG-1195 \citep{bond17,shvartzvald17}. 
The shortest duration event with a planetary mass ratio companion was MOA-2011-BLG-262 \citep{2014ApJ...785..155B}. 
Only a few microlensing planets with $t_{\rm E} \lesssim 10$ days have been found. 
This is largely because the detection efficiency of planets in short events is low \citep{2016ApJ...833..145S}. 
We need to find and analyze more events like MOA-2012-BLG-505, to obtain unbiased statistics of planets around low-mass hosts. \\

The microlensing method has the capability to determine the dependence of the distribution of exoplanets on the distance from the center of our Galaxy. 
The comparison of published planets by \citet{2016ApJ...830..150P} is not conclusive because of biases in the planetary sample and the lack of a proper detection efficiency calculation for this sample. 
Large statistical samples \citep{2016ApJ...833..145S} could be used for such a comparison if they are supplemented by lens mass measurements via microlensing parallax \citep{gaudi-ogle109,2014ApJ...785..155B,2011ApJ...741...22M} or host star brightness measurements \citep{bennett06,bennett15,batista15}. 
However, both of these methods are more effective for planetary systems in the disk. 
Space-based microlensing parallax measurements have somewhat of an advantage in this regard, although the Spitzer program is limited in its sensitivity to short duration events due to the necessity of identifying targets from ground-based data \citep{zhu17}. 
The Kepler K2 microlensing program is limited by its short duration \citep{2016PASP..128l4401H}. 
Prior to this paper, there have been five planets located with a $1 \sigma$ distance lower limit of $D_{\rm L} > 6.0$ kpc, i.e., likely to be in the Bulge. 
MOA-2011-BLG-028Lb \citep{2016ApJ...820....4S} was determined to be in the bulge due to an upper limit on the microlensing parallax, and MOA-2011-BLG-293Lb \citep{2014ApJ...780...54B} was determined to be in the bulge by detection of the lens star. 
MOA-2008-BLG-310Lb \citep{2010ApJ...711..731J}, OGLE-2015-BLG-0051Lb \citep{2016AJ....152...95H},  and MOA-2011-BLG-322Lb \citep{2014MNRAS.439..604S} were estimated to be in the bulge based on a Bayesian analysis that assumed that bulge and disk stars are equally likely to host planets, so these bulge planet identifications are less definitive than the first two. 
MOA-2011-BLG-028Lb, MOA-2008-BLG-310Lb and OGLE-2015-BLG-0051Lb are expected to be a Neptune-mass planet, Sub-Saturn mass planet and a sub-Jupiter mass planet in the Galactic bulge, respectively. 
MOA-2011-BLG-293Lb and MOA-2011-BLG-322Lb are super-Jupiter mass planets in the Galactic bulge. 
The MOA-2011-BLG-262L host has an unusually high $\mu_{\rm rel}$, which combined with the small $t_{\rm E}$ implies a low mass stellar host in the bulge or an apparently unbound planet hosting a Earth-mass moon. 
The Bayesian analysis for the OGLE-2016-BLG-1195 planetary system indicates a bulge lens \citep{bond17}, but the analysis of Spitzer data \citep{shvartzvald17} seems to indicate a microlensing parallax signal due to a star in the disk that is not rotating with the disk. 
The {\it priori} probability of microlensing parallax direction given its magnitude and the measured $\mu_{\rm rel}$ value is $\sim 0.003$. 
This suggests that the parallax measurement may have been compromised by a companion to the source or lens, so the nature of this event is as yet undetermined. 
Due to these ambiguities with MOA-2011-BLG-262Lb and OGLE-2016-BLG-1195Lb,  MOA-2012-BLG-505Lb is the sixth planet which is likely to be in the Bulge. 
However, the possibility that it is a planet orbiting a brown dwarf in the disk cannot be ruled out. 
Nevertheless, this discovery contributes to the statistics of planet distribution in our Galaxy. \\

We used the Bayesian analysis to estimate the probability distribution of the lens physical parameters because we couldn't measure the microlensing parallax. 
It is difficult to observe the microlensing parallax in the Bulge lens events because they are relatively short and their projected Einstein radius on the observer plane, $\tilde{r}_{\rm E}$ is large relative to 1 AU. 
$\tilde{r}_{\rm E}$ for the Bulge lens event is $3 \sim 10 \rm AU$ typically. 
It is difficult to detect the microlensing parallax for such events from the space telescope located at L2, however we may be able to detect these from the telescope located on $\sim$ AU away from the Earth like Spitzer \citep{2016ApJ...819...93S} and/or Kepler \citep{2016PASP..128l4401H}. 
If these instruments observe many events similar to MOA-2012-BLG-505, they will be able to determine the planet distribution in the Galaxy down to low-mass hosts. 
NASA's WFIRST satellite will increase this sample size significantly and reveal the Galactic distribution of exoplanets \citep{2015arXiv150303757S}. \\

We may identify the lens by future high resolution imaging with space telescope or ground base AO observation if it is a relatively massive star, or we may set a tight upper limit when it is really low-mass star.
Figure 7 shows the distributions of $I$, $J$, $H$, and $K$-band magnitudes for the lens star with extinction. 
The lens magnitudes are estimated from \citet{1995ApJS..101..117K} and \citet{2003A&A...402..701B}. 
We selected the isochrone model for a 5 Gyr brown-dwarf from \citet{2003A&A...402..701B}. 
The extinctions are estimated by using \citet{1989ApJ...345..245C}. 
The distributions look bimodal because the luminosity changes sharply around $0.07 M_{\rm \odot}$, i.e., at the boundary between brown-dwarf and main sequence stars, in the mass-luminosity relation. 
We cannot detect the lens star if it is brown-dwarf. 
However if the lens star is a hydrogen burning star, probably we will be able to detect it with JWST \citep{2006SSRv..123..485G} when it is separated from the source star after 10 years. \\

TS acknowledges the financial support from the JSPS, JSPS23103002, JSPS24253004 and JSPS26247023. The MOA project is supported by the grant JSPS25103508 and 23340064. 
Work by N.K. is supported by JSPS KAKENHI Grant Number JP15J01676. 
Work by A.F. is supported by JSPS KAKENHI Grant Number JP17H02871. 
NJR is a Royal Society of New Zealand Rutherford Discovery Fellow. 
AS is a University of Auckland Doctoral Scholar.

{}

\begin{figure}[htbp]
\begin{center}
  \includegraphics[scale=0.5,angle=-90]{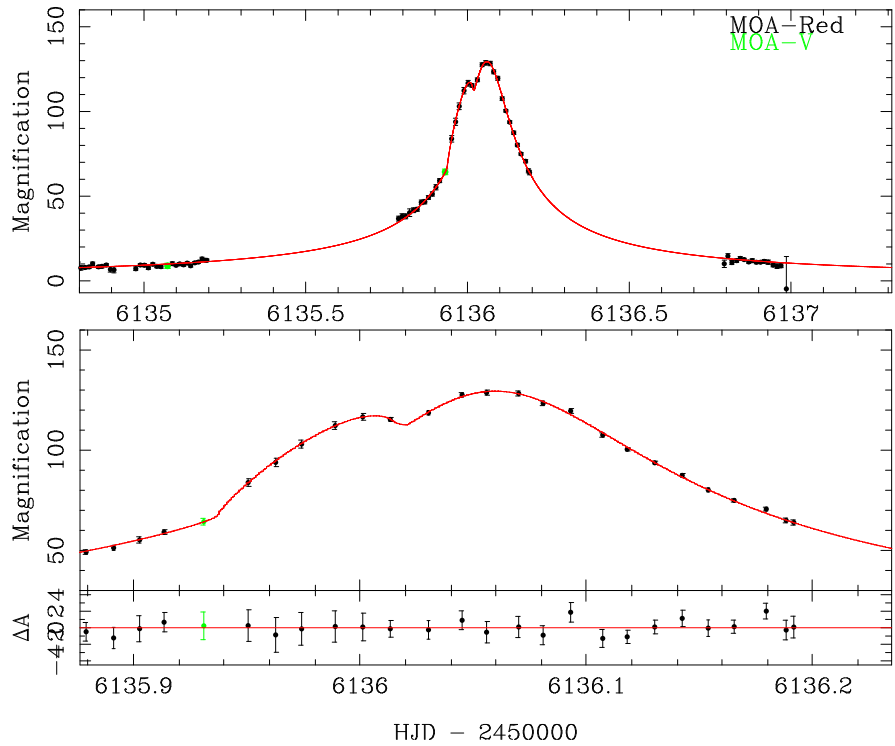}
  \caption{The light curve of event MOA-2012-BLG-505 with data from MOA-Red (black) and MOA-V (Green). The best-fit model is indicated by the red line. Middle and bottom panels show the detail of planetary signal and the residual from the best model respectively.}
\end{center}
\end{figure}

\begin{figure}[htbp]
\begin{center}
  \includegraphics[scale=0.5,angle=-90]{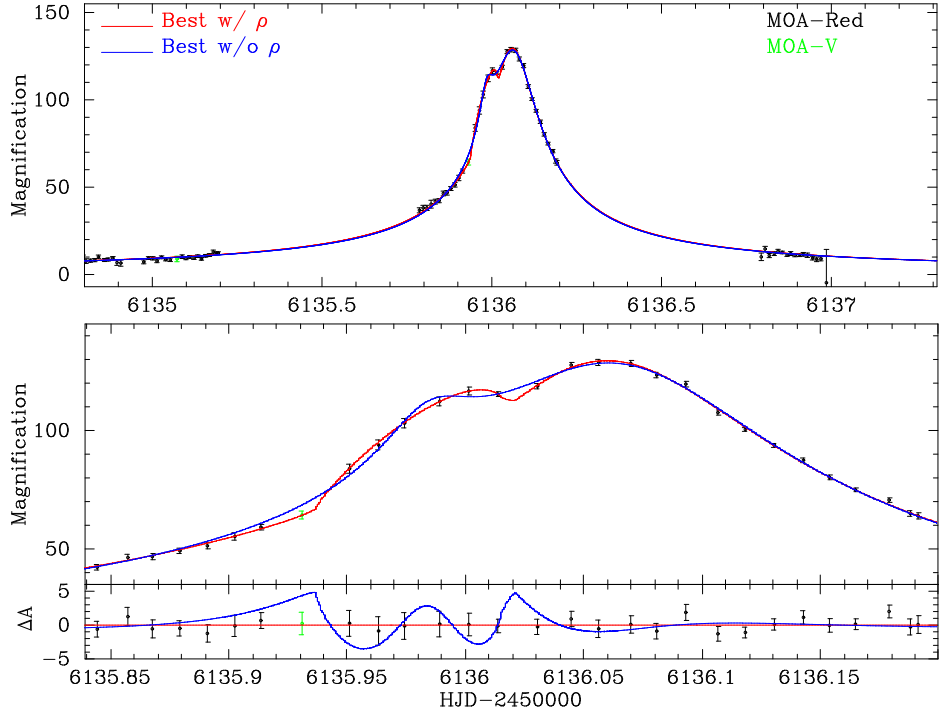}
  \caption{The best-fit model fitted with the finite source effect (red line) and the best-model fitted without the finite source effect (blue line). The $\chi^{2}$ of blue line is $\Delta \chi^{2} \sim 17.1$ larger than red one.}
\end{center}
\end{figure}

\begin{figure}[htbp]
  \begin{center}
    \begin{tabular}{c}
      \begin{minipage}{0.42\hsize}
        \begin{center}
          \includegraphics[scale=0.35,angle=-90]{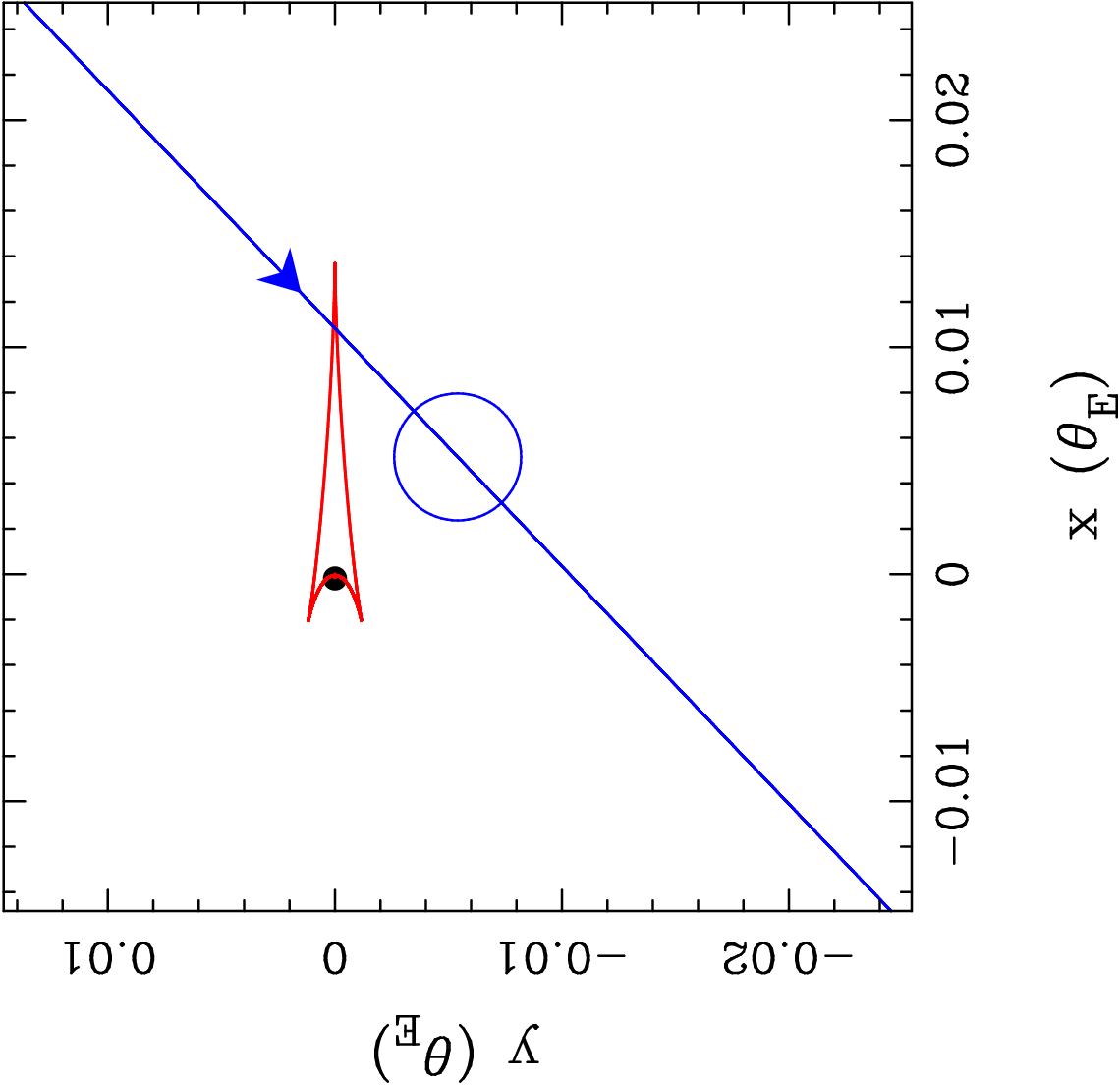}
        \end{center}
      \end{minipage}
      \begin{minipage}{0.42\hsize}
        \begin{center}
          \includegraphics[scale=0.35,angle=-90]{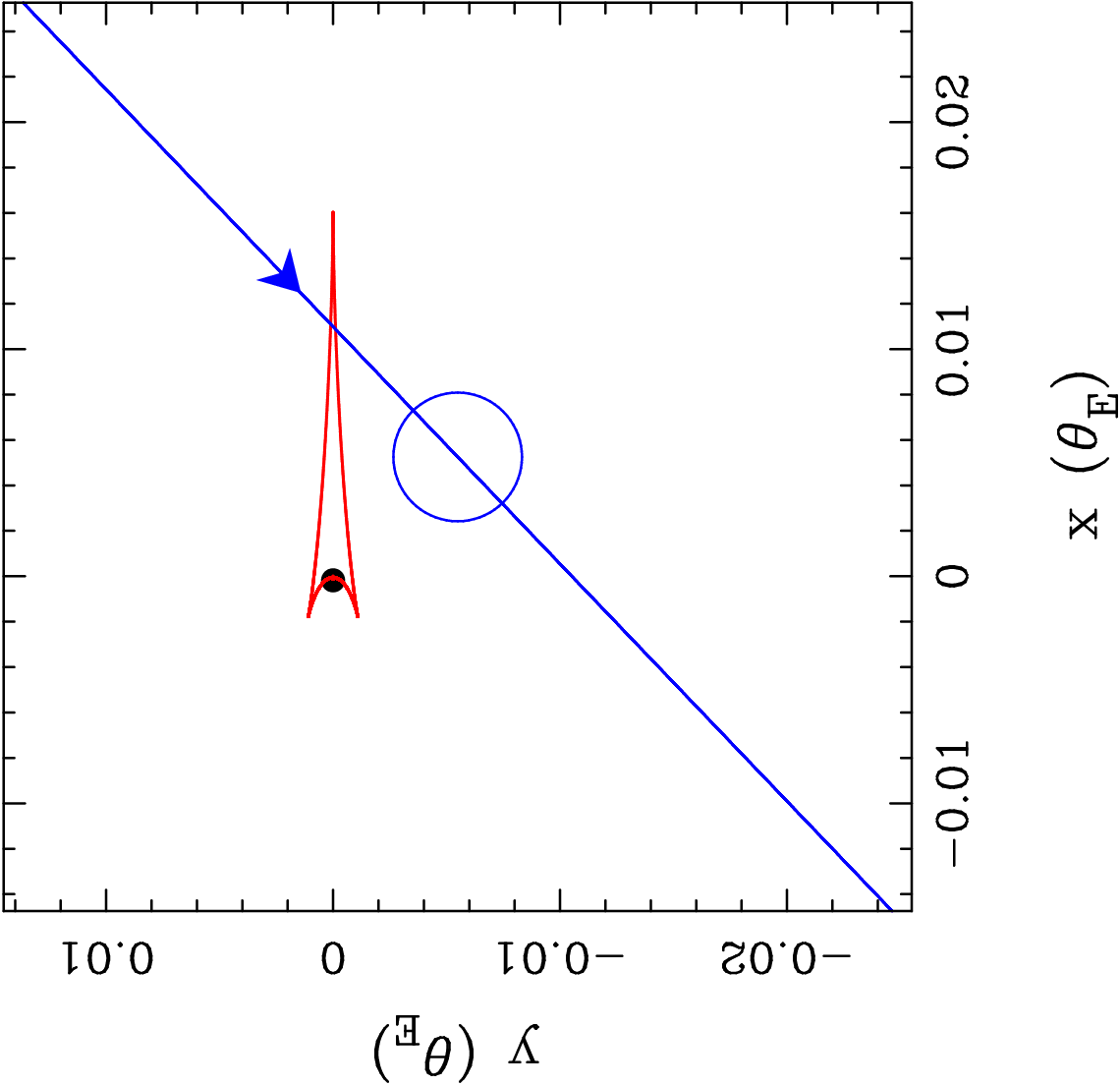}
        \end{center}
      \end{minipage}
    \end{tabular}
    \caption{Caustic geometries for the best-fit close (left) and wide (right) models indicated by the red curves. The blue line for each figures shows the source trajectory with respect to the lens system. The blue circle for each figures indicates the source star size.}
  \end{center}
\end{figure}

\begin{figure}[htbp]
\begin{center}
  \includegraphics[scale=0.5,angle=-90]{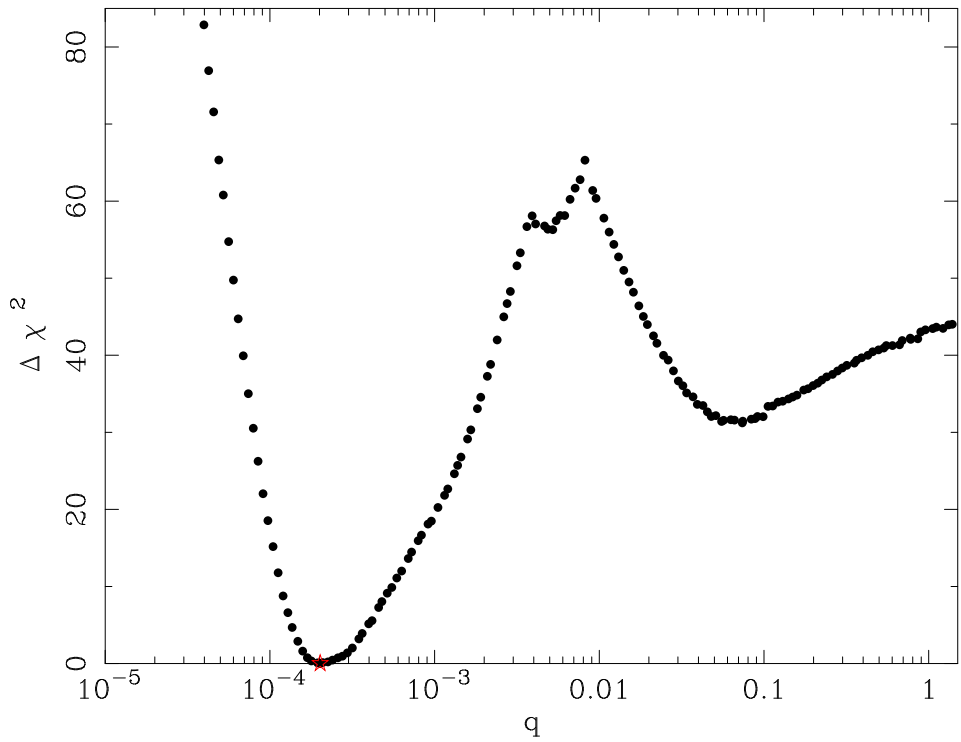}
  \caption{$\Delta \chi^{2}$ from the best-fit model as a function of $q$. The red star indicates the $\Delta \chi^{2}$ and $q$ values corresponding to the best fitting model. There is a local minimum at $q \sim 6.63 \times 10^{-2}$, and $\Delta \chi^{2}$ between the best and local minimum is about 31.}
\end{center}
\end{figure}

\begin{table}[htb]
\caption{The best-fit parameters and $1\sigma$ errors for close and wide models}
  \begin{center}
  \begin{tabular}{lccc}
\hline \hline
    parameter & units & close & wide \\
    error ($1\sigma$) &  & ($s<1$) & ($s>1$) \\ \hline
    $t_{0}$ & HJD - 2450000 & 6136.0557 & 6136.0558 \\
              &        & 0.0005 & 0.0005 \\
    $t_{E}$ & days & 10.0133 & 9.8335 \\
              &        & 1.0035 & 1.0506 \\
    $u_{0}$ & $10^{-3}$ & 7.4775 & 7.6002 \\
              &        & 0.8810 & 0.8568 \\
    $q$    & $10^{-4}$ & 2.0520 & 2.0521 \\
              &        & 0.5647 & 0.5528 \\
    $s$    &        & 0.8928 & 1.1266 \\
              &        & 0.0477 & 0.0653 \\
    $\alpha$ & radians & 2.3794 & 2.3782 \\
              &        & 0.0175 & 0.0160 \\
    $\rho$ & $10^{-3}$ & 2.7971 & 2.8322 \\
              &        & 0.3714 & 0.3703 \\
    $\theta_{\rm *}$ & ${\rm \mu as}$ & 0.342 & 0.342 \\
              &        & 0.052 & 0.052 \\
    $\theta_{\rm E}$ & mas & 0.122 & 0.121 \\
              &        & 0.025 & 0.024 \\
    $\mu_{\rm rel}$  & ${\rm mas \ yr^{-1}}$ & 4.3890 & 4.3971 \\
              &        & 0.7198 & 0.7196 \\
    $d.o.f.$ &  & 16004 & 16004 \\
    $\chi^{2}$ &  & 16010.859 & 16010.840 \\
\hline
  \end{tabular}
  \end{center}
\end{table}

\begin{figure}[htbp]
\begin{center}
  \includegraphics[scale=0.6,angle=-90]{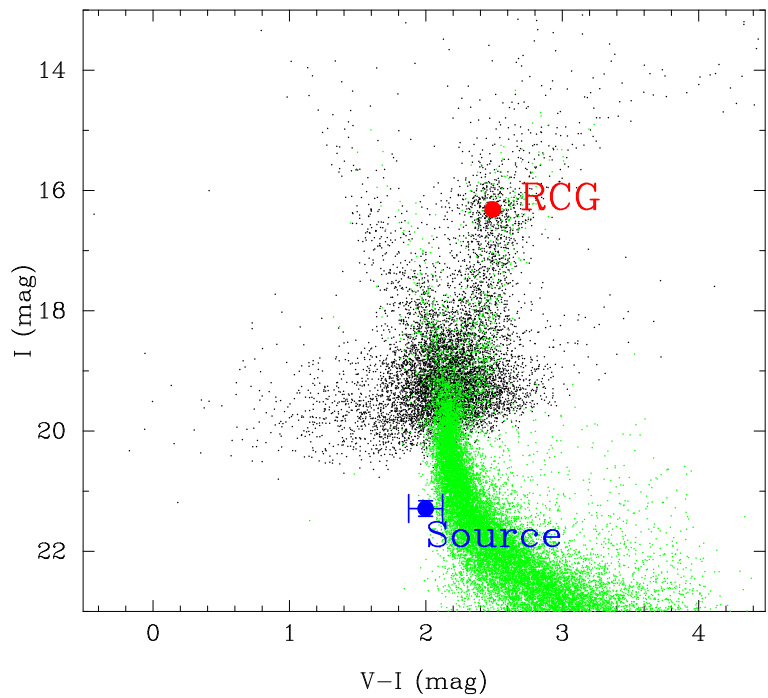}
  \caption{Color Magnitude Diagram (CMD) of OGLE-III stars within $2^{\prime}$ of MOA-2012-BLG-505 (black dots). The green dots show the HST CMD \citep{1998AJ....115.1946H}. The red point indicates the centroid of red clump giant, and the blue point indicates the source color and magnitude.}
\end{center}
\end{figure}

\begin{figure}[htbp]
\begin{center}
  \includegraphics[scale=0.45,angle=-90]{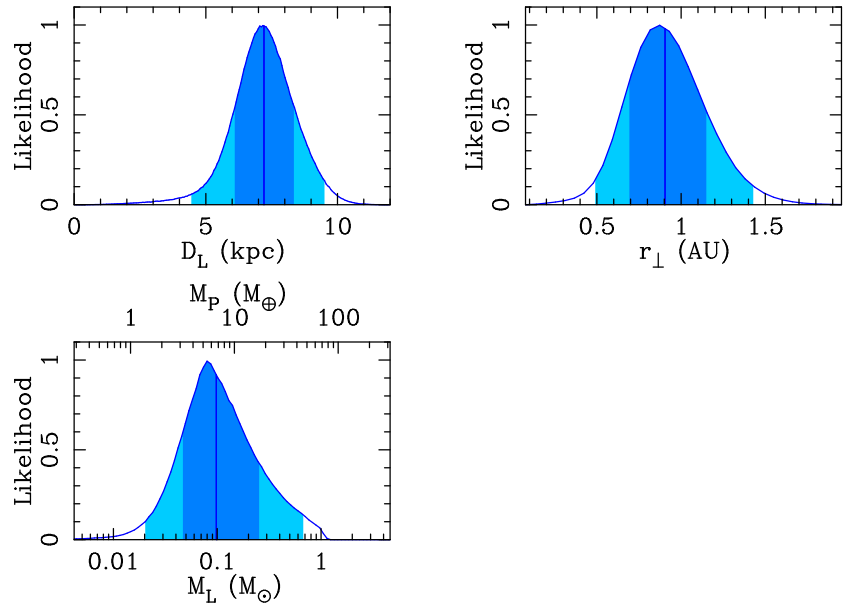}
  \caption{Relative probability distributions of lens system properties from our Bayesian analysis. The dark and light blue regions indicate the 68$\%$ and 95$\%$ confidence intervals respectively. The blue vertical lines indicate the median values of each of these distributions.}
\end{center}
\end{figure}

\begin{figure}[htbp]
\begin{center}
  \includegraphics[scale=0.5,angle=-90]{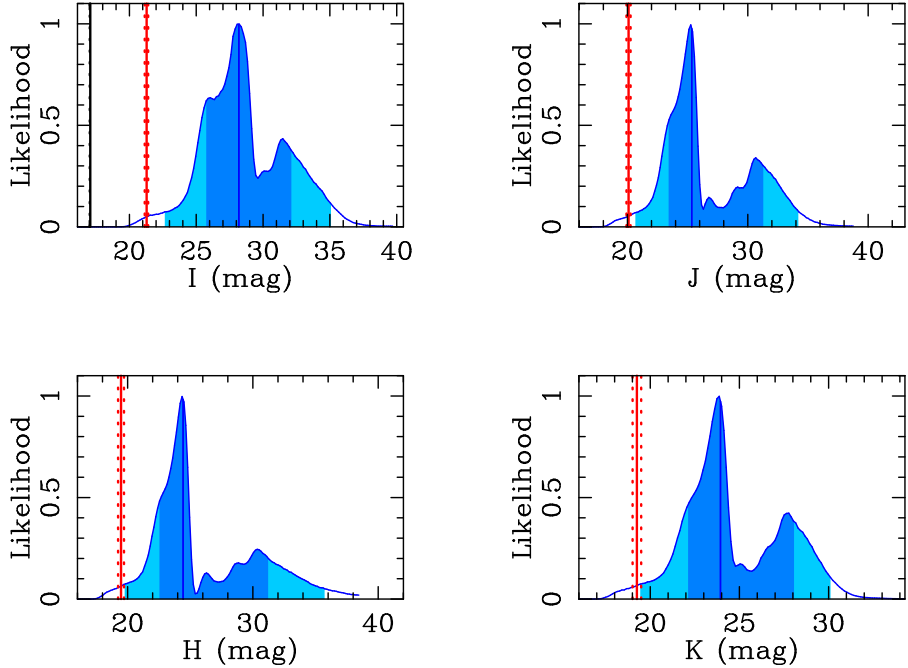}
  \caption{Relative probability distributions for the $I$, $J$, $H$ and $K$-band lens star magnitudes from our Bayesian analysis with extinction. The dark and light blue regions indicate the 68$\%$ and 95$\%$ confidence intervals respectively. The vertical blue lines indicate the median values of each of these distributions. The vertical red lines indicate the source star magnitudes for each bands. Red dotted lines are their $1\sigma$ errors. The source magnitudes are estimated from \citet{1995ApJS..101..117K} and \citet{2003A&A...402..701B}. Their extinctions are estimated from \citet{1989ApJ...345..245C}.}
\end{center}
\end{figure}

\begin{table}[htb]
\caption{The physical parameters and $1\sigma$ errors of MOA-2012-BLG-505}
  \begin{center}
  \begin{tabular}{lccc}
\hline \hline
    parameter & units & value & error($1\sigma$) \\ \hline
    $D_{\rm L}$ & kpc & 7.21 & $^{+1.14}_{-1.11}$ \\
    $M_{\rm L}$ & $M_{\odot}$ & 0.10 & $^{+0.16}_{-0.05}$ \\
    $m_{\rm p}$ & $M_{\oplus}$ & 6.70 & $^{+10.61}_{-3.51}$ \\
    $r_{\rm \perp}$ & AU & 0.90 & $^{+0.25}_{-0.21}$ \\
    $a$ & AU & 1.12 & $^{+0.67}_{-0.32}$ \\
    $\mu$ & mas\,yr$^{-1}$ & 4.72 & $^{+1.01}_{-0.91}$ \\
    $J$ & mag & 25.34 & $^{+5.94}_{-1.92}$ \\
    $H$ & mag & 24.43 & $^{+6.79}_{-1.89}$ \\
    $K$ & mag & 23.93 & $^{+4.13}_{-1.82}$ \\
\hline
  \end{tabular}
  \end{center}
\end{table}

\begin{figure}[htbp]
\begin{center}
  \includegraphics[scale=0.5,angle=-90]{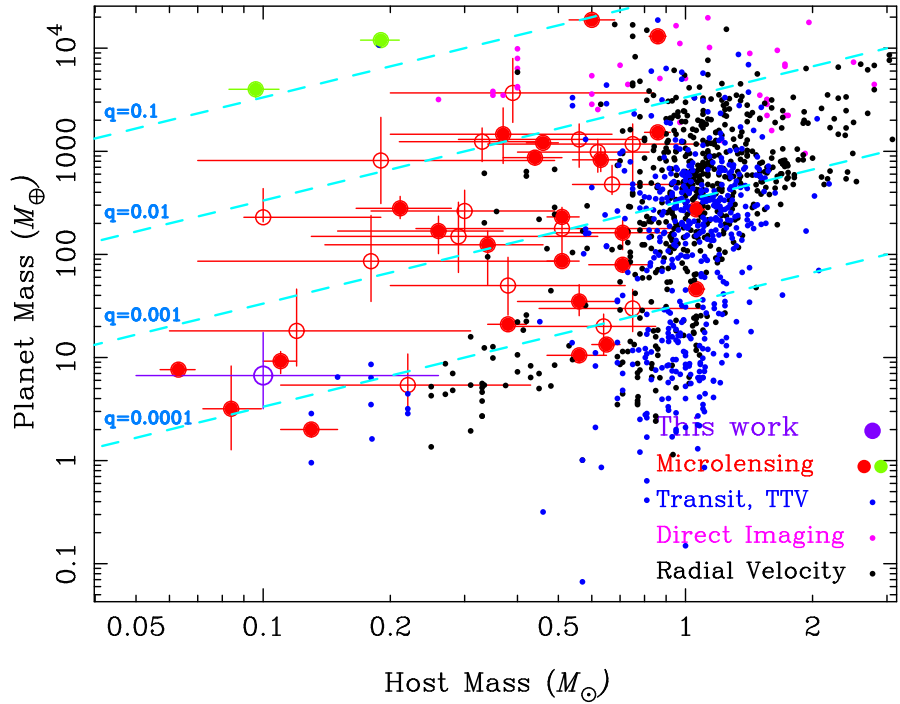}
  \caption{Distribution of exoplanets. The horizontal axis corresponds to the host mass and the vertical axis corresponds to the planet mass. The purple point indicates MOA-2012-BLG-505. The red, green, blue, magenta and black points indicate the planets found by Microlensing (planetary mass companion), Microlensing (binary mass companion), Transit \& TTV, Direct Imaging, and Radial Velocity, respectively. In Microlensing planets, filled circles indicates that their masses are measured and open circles indicate that their masses are estimated by a Bayesian analysis. The values of microlensing planets are from literature, while those of the others are from http://exoplanet.eu.}
\end{center}
\end{figure}

\newpage

\end{document}